\documentclass[11pt,a4paper]{article}
\usepackage{amsmath, amssymb, ulem, marvosym, wasysym, wrapfig}
\usepackage[dvipdfm]{graphicx}
\usepackage{wrapfig}

\def\un#1{\relax\ifmmode\@@underline#1\else
        $\@@underline{\hbox{#1}}$\relax\fi}

\let\du=\du                     

\def\a{\alpha}
\def\b{\beta}

\def\f{\phi}
\def\g{\gamma}

\def\m{\mu}
\def\n{\nu}

\def\L{\Lambda}

\def\bo{{\raise-.3ex\hbox{\large$\Box$}}}               
\def\TH{{\raise.2ex\hbox{$\displaystyle \bigodot$}\mskip-4.7mu \llap H \;}}
\def\face{{\raise.2ex\hbox{$\displaystyle \bigodot$}\mskip-2.2mu \llap {$\ddot
        \smile$}}}                                      

\def\abs#1{\left| #1\right|}                    
\def\leftrightarrowfill{$\mathsurround=0pt \mathord\leftarrow \mkern-6mu
        \cleaders\hbox{$\mkern-2mu \mathord- \mkern-2mu$}\hfill
        \mkern-6mu \mathord\rightarrow$}
\def\dvec#1{\vbox{\ialign{##\crcr
        \leftrightarrowfill\crcr\noalign{\kern-1pt\nointerlineskip}
        $\hfil\displaystyle{#1}\hfil$\crcr}}}           
\def\dt#1{{\buildrel {\hbox{\LARGE .}} \over {#1}}}     

\def\sfrac#1#2{{\vphantom1\smash{\lower.5ex\hbox{\small$#1$}}\over
        \vphantom1\smash{\raise.4ex\hbox{\small$#2$}}}} 
\def\bfrac#1#2{{\vphantom1\smash{\lower.5ex\hbox{$#1$}}\over
        \vphantom1\smash{\raise.3ex\hbox{$#2$}}}}       
\def\afrac#1#2{{\vphantom1\smash{\lower.5ex\hbox{$#1$}}\over#2}}    

\def\[{\lfloor{\hskip 0.35pt}\!\!\!\lceil}
\def\]{\rfloor{\hskip 0.35pt}\!\!\!\rceil}

\def\du#1#2{_{#1}{}^{#2}}

\def\ha{{\fracmm12}}

\def\un{\underline}
\def\fracmm#1#2{{{#1}\over{#2}}}

\def\low#1{{\raise -3pt\hbox{${\hskip 0.75pt}\!_{#1}$}}}

\def\Dot#1{\buildrel{_{_{\hskip 0.01in}\bullet}}\over{#1}}
\def\dt#1{\Dot{#1}}

\newskip\humongous \humongous=0pt plus 1000pt minus 1000pt

\newif\ifdtup

\newcommand{\be}{\begin{equation}}
\newcommand{\ee}{\end{equation}}
\newcommand{\nbe}{\begin{equation*}}
\newcommand{\nee}{\end{equation*}}

\newcommand{\lb}{\label}

\def\lessim{\lower0.6ex\hbox{$\,$\vbox{\offinterlineskip\hbox{$<$}\vskip1pt\hbox{$\sim$}}$\,$}}
\def\grtsim{\lower0.6ex\hbox{$\,$\vbox{\offinterlineskip\hbox{$>$}\vskip1pt\hbox{$\sim$}}$\,$}}

\numberwithin{equation}{section}

\begin{document}

\begin{titlepage}

\begin{center}

October 2014 \hfill IPMU14-0318\\

\noindent
\vskip2.0cm
{\Large \bf 

The $f(R)$ Gravity Function of the Linde Quintessence 

}

\vglue.3in

{\large
Sergei V. Ketov~${}^{a,b,c}$ and Natsuki Watanabe~${}^a$ 
}

\vglue.1in

{
${}^a$~Department of Physics, Tokyo Metropolitan University \\
Minami-ohsawa 1-1, Hachioji-shi, Tokyo 192-0397, Japan \\
${}^b$~Kavli Institute for the Physics and Mathematics of the Universe (IPMU)
\\The University of Tokyo, Chiba 277-8568, Japan \\
${}^c$~Institute of Physics and Technology, Tomsk Polytechnic University\\
30 Lenin Ave.,Tomsk 634050, Russian Federation \\
}

\vglue.1in
ketov@tmu.ac.jp, watanabe-natsuki1@tmu.ac.jp

\end{center}

\vglue.3in

\begin{center}
{\large\bf Abstract}
\end{center}

We calculate the $f(R)$ gravity function in the dual gravity description of the quintessence model with a quadratic (Linde)
scalar potential and a positive cosmological constant. We find that in the large curvature regime relevant to chaotic inflation
in early Universe, the dual $f(R)$ gravity is well approximated by the (matter) loop-corrected Starobinsky inflationary model. In the small curvature regime relevant to dark energy in the present Universe, the $f(R)$ gravity function reduces to the Einstein-Hilbert
one with a positive cosmological constant.

\end{titlepage}


\section{Introduction}\label{sec:Intro}

Theoretical models of cosmological inflation (or primordial dark energy) in Early Universe and those of dynamical dark energy (in the Present Universe) are known to be easily constructed by the use of modified
$f(R)$ gravity or quintessence. The standard treatment usually includes the remarkable duality between an
 $f(R)$-gravity model and the classically equivalent scalar-tensor gravity (quintessence) model 
 by the Legendre-Weyl transform from the Jordan frame to the Einstein frame, with the standard (quintessence) cosmology in terms of the dual (inflaton)  scalar potential (see Sec.~2 or reviews 
 \cite{svrev,tsu,norev,clrev,myrev}  for details). 

The most economical, simple and viable inflationary model on the $f(R)$ gravity side is given by the 
Starobinsky model  \cite{star1,star2,mchi,star3}, with an action
\be \lb{stara}
S[g] = \int \mathrm{d}^4x\sqrt{-g} \left[ -\ha R +\frac{1}{12M^2}R^2\right]~,
\ee
in terms of 4D spacetime metric $g_{\m\n}(x)$ of the scalar curvature $R$. We use the natural units
with the reduced Planck mass $M_{\rm Pl}=1$, unless is otherwise stated. Slow-roll inflation takes place in the high-curvature regime  (with $M_{\rm Pl}\gg H \gg M$ and $|\dt{H}|\ll H^2$), where the second term in 
Eq.~(\ref{stara}) dominates. The inflationary model (\ref{stara}) has the only mass parameter $M$ that is fixed by the observational Cosmic Microwave Background (CMB) data as $M=(3.0 \times10^{-6})(\fracmm{50}{N_e})$ where $N_e$ is the e-foldings number. The predictions of the Starobinsky model for the spectral indices  $n_s\approx 1-2/N_e\approx 0.964$, $r\approx 12/N^2_e\approx0.004$ and low non-Gaussianity are in agreement with the WMAP and PLANCK data ($r<0.13$ and $r<0.11$, respectively, at 95\% CL) \cite{planck2}, but are in apparent {\it disagreement} with the BICEP2 measurements ($r=0.2+0.07,-0.05$) \cite{bicep2}, even after the dust contribution adjustment \cite{bicep2c}.

The action (\ref{stara}) can be dualized by the Legendre-Weyl transform \cite{lwtr,kkw}  to the standard (quintessence)  action of the Einstein gravity coupled to a single (canonically normalized) physical scalar (inflaton or scalaron) $\f$ having the scalar potential
\be \lb{starp}
V(\f) = \fracmm{3}{4} M^2\left( 1- e^{-\sqrt{\frac{2}{3}}\f }\right)^2~.
\ee
The exit from the Starobinsky inflation goes to a Minkowski vacuum, though a small positive cosmological constant can always be added in order to shift the Minkowski vacuum to a de-Sitter vacuum that does not affect the inflation.

On the quintessence side, the simplest inflationary model  with a {\it quadratic} scalar potential was proposed by Linde \cite{linde}. It predicts $r\approx 8/N_e=0.16\left(\fracmm{50}{N_e}\right)$ in good 
agreement with the BICEP2 data \cite{bicep2}. By adding a small cosmological constant to the Linde
scalar potential one can also take into account the present dark energy (after an exit from the Linde inflation 
and inflaton decay).

To the best of our knowledge, the $f(R)$ gravity function for the Linde scalar potential is unknown.   The main purpose of this letter is to fill this gap in the literature.  We calculate that function and find it to be
non-trivial, being related to the loop-corrected Starobinsky model (Sec.~4).

Our paper is organized as follows. In Sec.~2 we review the Legendre-Weyl transform and give its inverse form. In Sec.~3 we find the exact $f(R)$ gravity function for the Linde scalar potential in a parametric form, and analytically study its limits in the cases of high and low spacetime scalar curvature, respectively.
Sec.~4 is our conclusion.    

Throughout this paper we use the natural units, $c = \hbar = 1$, and the space-time signature 
$\eta_{\mu \nu} = {\rm diag} (1, - 1, - 1, - 1)$. The Einstein-Hilbert action with a cosmological constant $\L$ reads
\begin{equation} \lb{eh}
 S_{\rm EH} =  -\frac{1}{2 \kappa^{2}} \int d^{4}x \sqrt{- g}~(R + 2 \Lambda),
 \end{equation} 
where $R$ is the scalar curvature, $\kappa^{2} = \fracmm{1}{M_{\rm Pl}^{2}} = 1.7 \times 
10^{- 37}~{\rm GeV^{-2}}$, and $M_{\rm Pl} = ( 8 \pi G_{\rm N} )^{- 1 / 2}$ is the (reduced) Planck mass in terms
of the Newton constant $G_{\rm N}$. In our notation, the cosmological constant $\L$ is positive and the scalar
curvature is negative in a {\it de-Sitter} (dS) spacetime (like the Present Universe), and vice versa 
in an Anti-de-Sitter (AdS) spacetime.

\section{The Legendre-Weyl transform and its inverse}\label{sec:Setup}

An $f(R)$ gravity action 
\begin{equation} \lb{fg}
  S_{f} = \frac{1}{2 \kappa^{2}} \int d^{4}x \sqrt{- g}~ f(R)
\end{equation} 
subject to the classical and quantum stability conditions (or, equivalently, no ghosts and tachyons, respectively)
 \cite{svrev,tsu,norev,clrev,myrev}
 \begin{equation} \lb{stab2}
  f'(R) <0 \qquad {\rm and} \qquad f''(R) >0 ~~,
\end{equation} 
where the primes denote differentiations,  is classically equivalent to 
\begin{equation} \lb{eq}
 S[g_{\m\n},\chi] = \frac{1}{2 \kappa^{2}} \int d^{4}x \sqrt{- g}~ \left[ f'(\chi) (R - \chi) + f(\chi) \right]
\end{equation}
with the real scalar field $\chi$, provided that $f''\neq 0$ that we always assume. 
The equivalence can be easily verified because the $\chi$-field equation implies $\chi=R$. 
The first condition (\ref{stab2}) also guarantees the existence of the dual (quintessence) description.

The factor $f'$ in front of the $R$ in eq.~(\ref{eq}) can be eliminated by a Weyl transformation of 
metric $g_{\m\n}$, so that one can transform the action (\ref{eq}) into the action of the scalar field 
$\chi$ miminally coupled to the Einstein gravity and having the scalar potential 
\be \lb{spot}
 V = \dfrac{\chi f'(\chi) - f(\chi)}{2 \kappa^{2} f'(\chi)^{2}}~~.
\ee
The kinetic term of $\chi$ becomes canonically normalized after the field refedinition
\begin{equation} \lb{fred}
  f'(\chi) = - \exp \left( - \sqrt{\frac{2}{3}} \kappa \phi \right)
\end{equation}
in terms of the new scalar field $\phi$. As a result, the action $S[g_{\m\n},\chi(\phi)]$  takes the standard
quintessence form.

Differentiating the scalar potential $V$ in Eq.~(\ref{spot}) with respect to $\phi$ yields 
\begin{equation} \lb{diff1}
    \frac{d V}{d \phi}   = \frac{d V}{d \chi} \frac{d \chi}{d \phi} 
      = \frac{1}{2 \kappa^{2}} \left[ \frac{\chi f'' + f' - f'}{f'^{2}} - 2 \frac{\chi f' - f}{f'^{3}} f'' \right] 
\frac{d \chi}{d \phi}~~,
\end{equation}
where we have
\begin{equation} \lb{diff2}
  \frac{d \chi}{d \phi}
    = \frac{d \chi}{d f'} \frac{ d f'}{d \phi}
    = \frac{d f'}{d \phi} \left/ \frac{d f'}{d\chi} \right.
    = - \sqrt{\frac{2}{3}} \kappa \frac{f'}{f''}~~.
\end{equation}
It implies that
\begin{equation} \lb{derv}
  \frac{d V}{d \phi} = \frac{\chi f' - 2 f}{\sqrt{6} \kappa f'^{2}}~~.
\end{equation}

Combining Eqs.~(\ref{spot}) and (\ref{derv})  yields $R$ and $f$ in terms of the scalar potential $V$ as follows:
\begin{align} \lb{inv}
  & R = - \left( - \sqrt{6} \kappa \frac{d V}{d \phi} + 4 \kappa^{2} V \right) \exp \left( - \sqrt{\frac{2}{3}} \kappa \phi \right),  \\
  & f = \left( - \sqrt{6} \kappa \frac{d V}{d \phi} + 2 \kappa^{2} V \right) \exp \left( - 2 \sqrt{\frac{2}{3}} \kappa \phi \right).
\end{align}
These two equations define the function $f(R)$ in the parametric form, in terms of a given scalar potential $V(\phi)$. 

In the case of the Higgs-like (or, more precisely, the uplifted $W$-shape) scalar potential for the present dark energy,
 the corresponding $f(R)$ gravity function was found in Ref.~\cite{kwl}.
 
\section{The $f(R)$ gravity dual of the Linde quintessence}\label{sec:Linde}

We are now in a position to compute the $f(R)$ gravity function for the Linde scalar potential of a canonically normalized inflaton $\f$
 \cite{linde}, 
\begin{equation}  \label{Lpot}
  V_{\rm L}(\f) = \frac{m^{2}}{2} \phi^{2} + V_{0}~,
\end{equation}
in order to get its dual (equivalent) gravitational description. The first term in Eq.~(\ref{Lpot}) is supposed to
dominate during chaotic inflation in early Universe, whereas the second term (cosmological constant) is supposed 
to dominate in the Present Universe (dark energy), well after the end of inflation followed by inflaton decay (reheating).
Accordingly, the parameters of Eq.~(\ref{Lpot}) have to be fixed by current observations as 
\begin{equation} \label{pval}
  m \approx 6 \times 10^{-6} \qquad {\rm and} \qquad
  V_{0} \approx 10^{-120}~~.
\end{equation}
It is worth mentioning here that the Linde inflation is consistent with the relatively high (vs. that of the Starobinsky model)
tensor-to-scalar ratio $r$, which is measurable via a detection of the B-mode polarization of the CMB radiation.

\begin{figure}[ht]
  \begin{minipage}{0.5\hsize}
    \centering
      \includegraphics[width=\hsize]{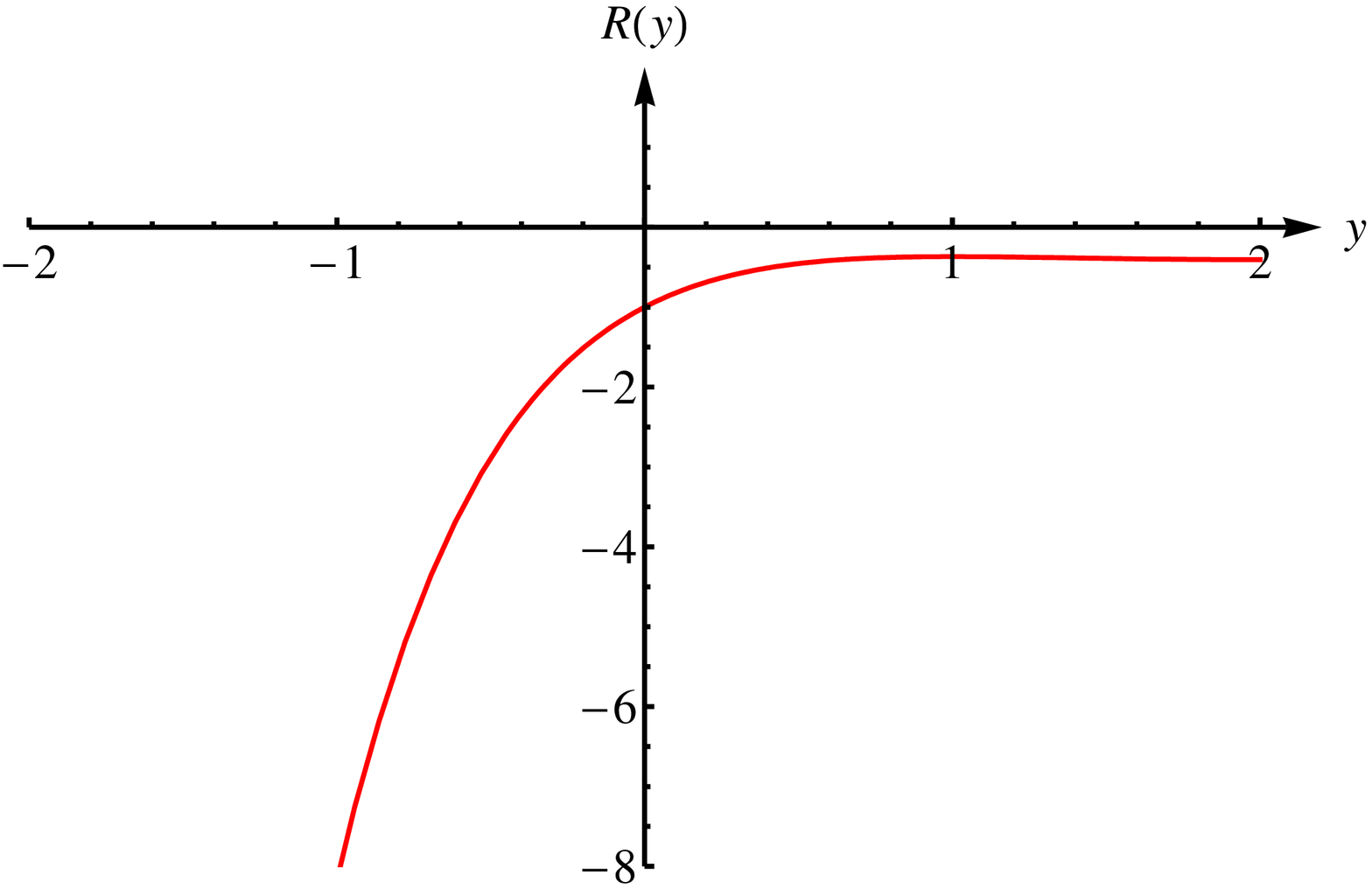}
      \caption{\small The case of $V_{0} > 0$.}
        \label{posi}
  \end{minipage}
  \begin{minipage}{0.5\hsize}
    \centering
      \includegraphics[width=\hsize]{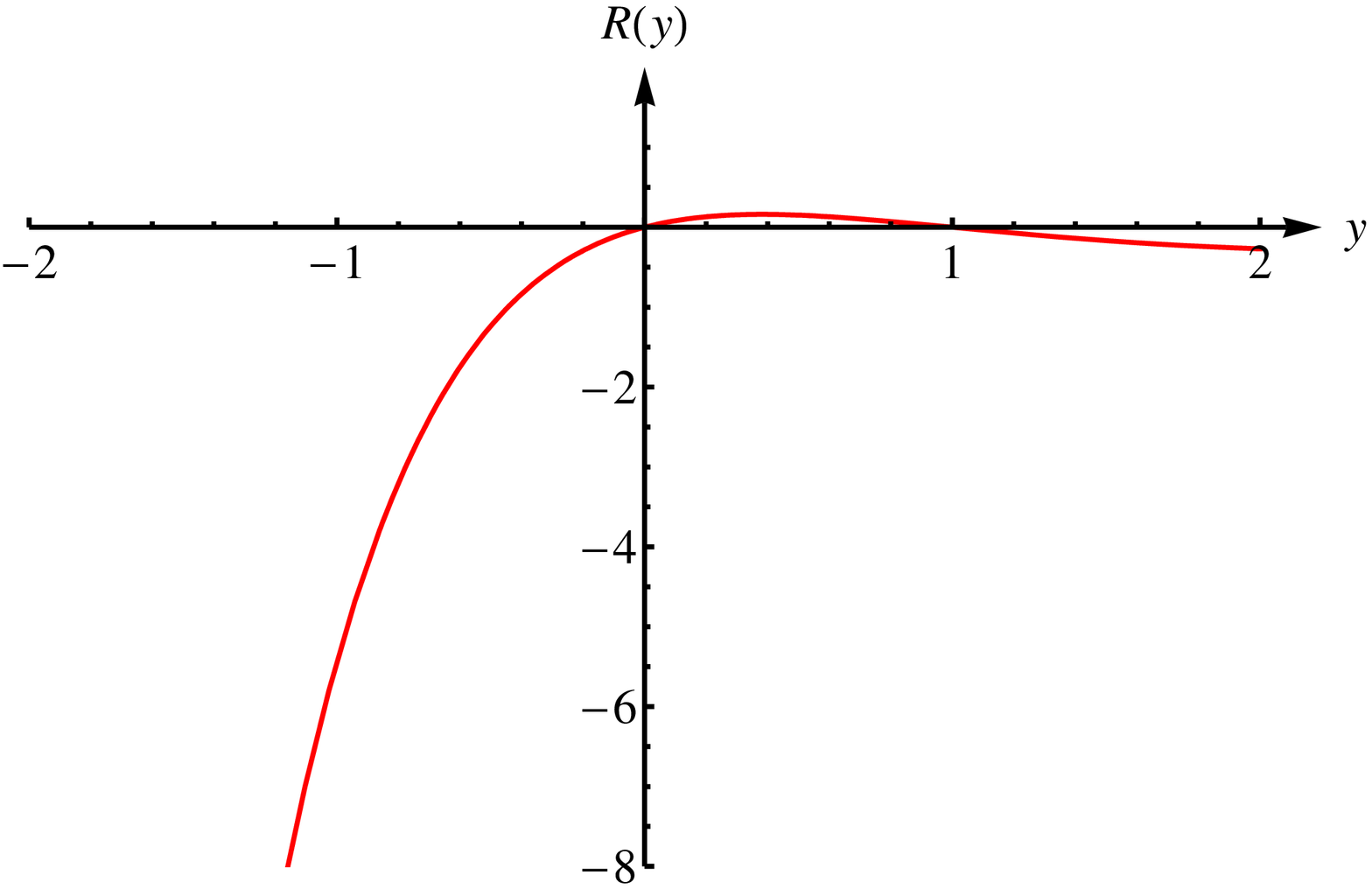}
      \caption{\small The case of $V_{0} = 0$.}
        \label{zero}
  \end{minipage}
\end{figure}

In the case of Eq.~(\ref{Lpot}), the inverse transform of Sec.~2 yields
\begin{align}
  & R = - 3 m^{2} \left( y^{2} - y + \frac{4 V_{0}}{3 m^{2}} \right) e^{-y}~~,  \label{Ry1} \\
  & f = \frac{3}{2} m^{2} \left( y^{2} - 2 y + \frac{4 V_{0}}{m^{2}} \right) e^{-2y}~~, \label{fy1}
\end{align}
where we have rescaled the inflaton field as $y = \sqrt{\frac{2}{3}}~ \phi$. These equations give the exact solution to the
$f(R)$ gravity function of the Linde quintessence in the parametric form. Figures (\ref{posi}) and (\ref{zero})  
show the behavior of the function $R(y)$ with a positive $V_0$ and the vanishing $V_{0}$, respectively, where 
the factor $3 m^{2}$ is absorbed into the normalization of $R$ and $f$.

In the {\it large} curvature (or large field) approximation relevant to inflation, Eq.~\eqref{Ry1} can be greatly simplified and solved as 
\begin{equation}
  R \approx - 3 m^{2} e^{-y}
  \qquad  {\rm and}\qquad 
   y(R) \approx - \ln \frac{-R}{3 m^{2}}~~.
\end{equation}
Substituting the $y(R)$ into Eq.~\eqref{fy1} yields
\begin{equation}  \label{fL1}
  f^{\rm E}(R) = \frac{3}{2} m^{2} \left( \frac{R}{3m^{2}} \right)^{2}
    \left[ \left( \ln \frac{|R|}{3 m^{2}} \right)^{2} + 2 \ln \frac{|R|}{3 m^{2}} + \frac{4 V_{0}}{3 m^{2}} \right]~~,
\end{equation}
where the superscript E refers to the Early Universe. During (slow-roll) inflation, the $R^2$ term dominates over the $R$ term in
Eq.~(\ref{stara}), so that the $f(R)$ gravity function (\ref{fL1}) is similar to the Starobinsky function proportional to $R^2$, though being corrected by the logarithmic terms (a cosmological constant can be safely ignored during an early universe inflation). 

It is not difficult to verify the stability conditions (\ref{stab2}) during inflation, when using our result  (\ref{fL1}).
We always assume that the scale of inflation is well below the Planck scale. We find the conditions
\begin{equation}  \label{Lcsc}
     2m^2 < \abs{R} \ll 1~~,
\end{equation}
which mean
\begin{equation} \label{Lcsc2}
7.2 \times 10^{-11}  < \abs{\frac{R}{M_{\rm Pl}^{2}}} \ll 1~~,
\end{equation}
where we have used  Eq.~(\ref{pval}) and have ignored the $V_0$ contribution.

The inflationary function $f^{\rm E}(R)$ can always be  represented in the Starobinsky-type form
\begin{equation} \lb{starfo}
  f^{\rm E}(R) = R^{2} A(R)
\end{equation}
in terms of another (positive) function $A(R)$. A slow-roll inflation of the Starobinsky-type can be achieved by
demanding the function $A(R)$ to be ``slowly varying" in the sense \cite{myrev}
\begin{align}
    & |A'(R)| \ll \frac{A(R)}{|R|}  \label{ARcond} ~~,\\
    & |A''(R)| \ll \frac{A(R)}{R^{2}}~~. \label{ARRcond}
\end{align}

In our case  (\ref{fL1}) we have 
\begin{equation}
  A(R) = \frac{1}{6 m^{2}} \ln \frac{|R|}{3 m^{2}} \left( \ln \frac{|R|}{3 m^{2}} + 2 \right)~.
\end{equation}
We find that the conditions  (\ref{ARcond}) and (3.11) imply 
\begin{equation}  \label{Lcsc3}
     \abs{R} \gg 3 m^2
\end{equation}
or
\begin{equation}
\abs{\frac{R}{M_{\rm Pl}^{2}}} \gg 1.1\times 10^{-10}~~,
\end{equation}
where we have used  Eq.~(\ref{pval}) and have ignored the $V_0$ contribution again.
Hence, the conditions (\ref{ARcond}) and (3.11) are consistent with those of Eq.~(\ref{Lcsc2}).

Similarly, in the {\it small} curvature (or small field) approximation, we have
\begin{equation} \lb{scap}
  R = 3 m^{2} \left( y - \frac{4 V_{0}}{3 m^{2}} \right) + \mathcal{O}(y^{2},~V_{0} y)
  \qquad {\rm and} \qquad  y \approx \frac{R + 4 V_{0}}{3 m^{2}}~~.
\end{equation}
Substituting them into Eq.~\eqref{fy1} gives rise to the $f(R)$ gravity function
\begin{equation} \lb{ehcc}
  f^{\rm P}(R) = \frac{1}{6 m^{2}} \left[ R^{2} - 2 ( 3 m^{2} - 4 V_{0} ) R - 4 V_{0} ( 3 m^{2} - 4 V_{0} ) \right]
    e^{-2 \frac{R + 4 V_{0}}{3 m^{2}}}~~,
\end{equation}
where  the superscript P refers to the Present Universe. After dropping the terms beyond the first order in $R$ and $V_{0}$,  
we arrive at the Einstein-Hilbert action with a cosmological constant $V_0$, as it should, namely,
\begin{equation}  \label{fS1}
  f^{\rm P}(R) = - R - 2 V_{0} + \mathcal{O}(R^{2},~V_{0}^{2},~V_{0} R)~~.
\end{equation}

\begin{figure}[ht]
  \begin{minipage}{0.45\hsize}
    \centering
      \includegraphics[width=\hsize]{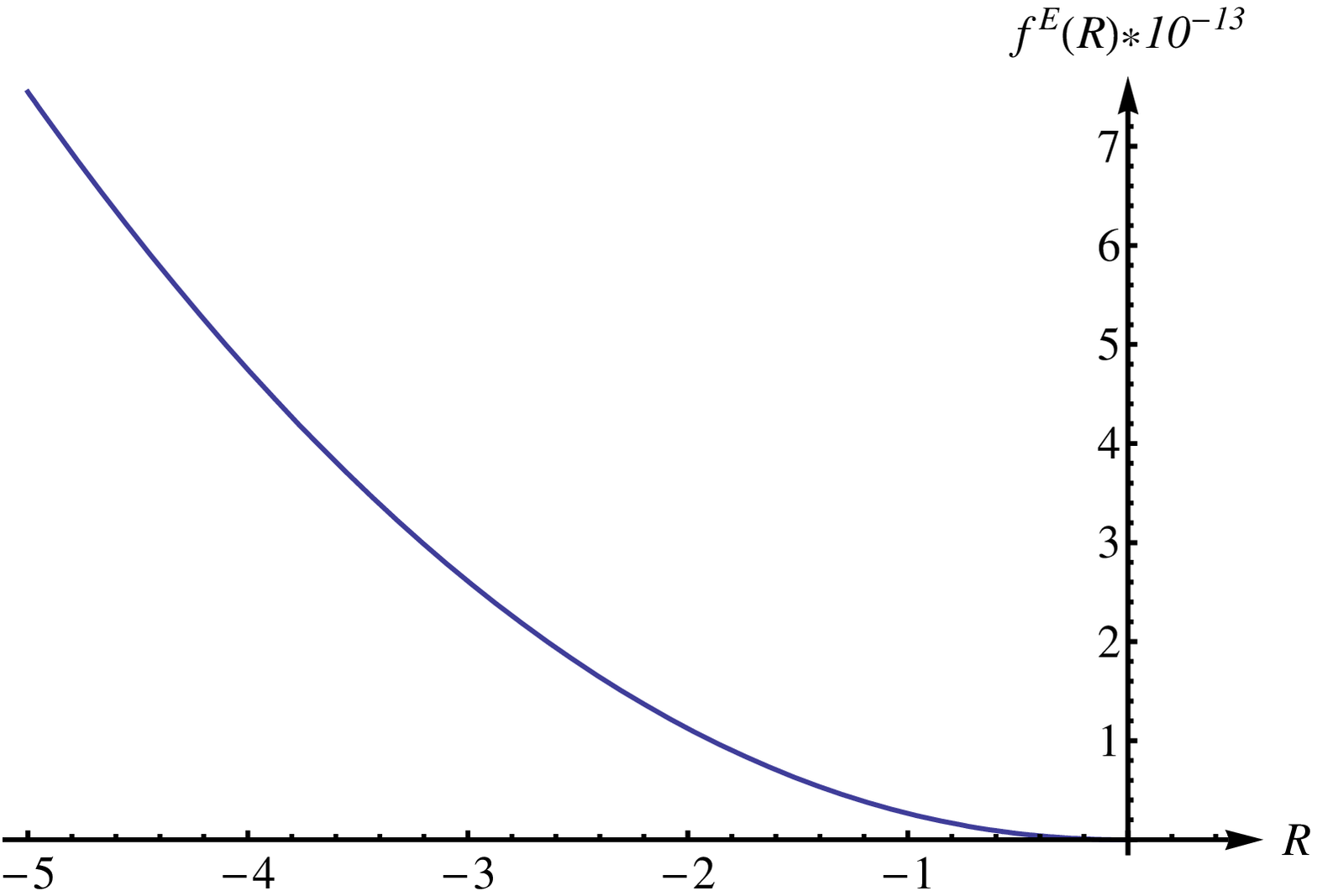}
      \caption{\small The function $f^{\rm E}(R)$.}
        \label{fe}
  \end{minipage}
  \begin{minipage}{0.5\hsize}
    \centering
      \includegraphics[width=\hsize]{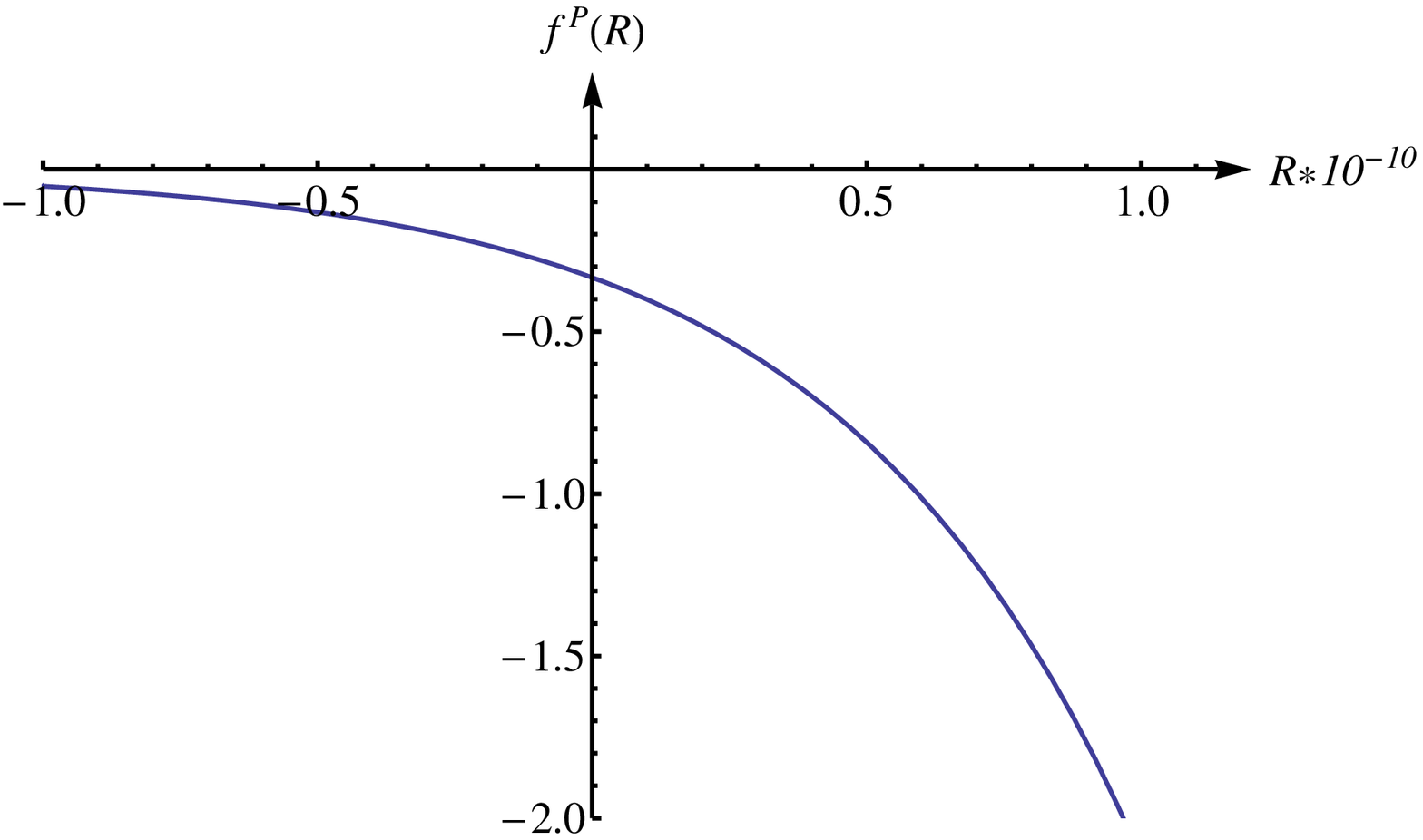}
      \caption{\small The function $f^{\rm P}(R)$.}
        \label{fp}
  \end{minipage}
\end{figure}

The profiles of the functions $f^{\rm E}(R)$ and $f^{\rm P}(R)$ are given in Figs.~(\ref{fe}) and (\ref{fp}) with the rescaled
argument $R$.

Checking the stability conditions (\ref{stab2}) in the low curvature approximation with the $f(R)$ gravity 
function (\ref{ehcc}) results in the condition
\begin{equation}  \label{range2}
     R <  1.1 m^2~~,
\end{equation}
where we have used  Eq.~(\ref{pval}) and have ignored the $V_0$ contribution too.  It implies
\begin{equation}
\frac{R}{M_{\rm Pl}^{2}} < 4 \times 10^{-11}
\end{equation}
that is obviously satisfied in our Present Universe.

Of course, stability is {\it not} an issue in the case of the Linde scalar potential  (\ref{Lpot}) with $m^2>0$ and $V_0\geq 0$.
However, a stability analysis becomes non-trivial on the dual gravity side. The bounds on the scalar curvature found above
should be merely considered as the restrictions on the approximation to be used.

\section{Conclusion}

Our main new results are given by Eqs.~(3.3), (3.4), (3.6) and (3.16). We also verified the stability conditions and gave the profiles of the relevant $f(R)$ gravity functions.

It is remarkable that our result (\ref{fL1}) for the $f(R)$ gravity function of the Linde inflation in the large curvature regime takes the form of the quantum-corrected Starobinsky inflationary model, with the logarithms representing the {\it loop} corrections of matter fields in curved space-time, just in the original spirit of the Starobinsky inflation 
\cite{stargu}~\footnote{See also Ref.~\cite{oe} for some explicit examples of such quantum corrections.}.
For example, the multi-loop (renormalization-group-improved) {\it Ansatz} for such quantum corrections
was proposed in Ref.~\cite{mass} in the form
\begin{equation}  \label{mloop}
  f^{\rm M}(R) = \m^{-1} R^2\left[ 1+\g \ln \left( \fracmm{R^2}{\m^2}\right)\right]^{-1}  
\end{equation}
with the renormalization group scale $\m$ and the parameter $\g$. Its expansion in powers of $\g$ leads to all powers of the logarithm in the pre-factor of $R^2$, in all the loop-orders.

Equation (\ref{fL1}) is also to be compared with the {\it Ansatz} of the one-loop corrected Starobinsky function (in
our notation)
\begin{equation}  \label{1loop}
  f^{\rm D}(R) = - R +\a R^2 +\b R^2 \log (-R)
\end{equation}
with some coefficients $\a$ and $\b$, used in Ref.~\cite{ijtwz} for the purpose of enhancement of the tensor-to-scalar
ratio $r$ of the simplest Starobinsky model  (\ref{stara}) towards its matching with the BICEP2 data \cite{bicep2}. As was
demonstrated in Ref.~\cite{ijtwz}, despite the enhancement in $r$, the function (\ref{1loop}) is disfavored by the
BICEP2 measurements for any values of the coefficients $\a$ and $\b$. However, as is clear from a simple comparison of 
Eq.~(\ref{1loop}) with our Eq.~(\ref{fL1}), in order to reach a viable enhancement, one should also take into account the
quantum corrections given by  the logarithm {\it squared}, of the type $R^2\log^2(-R)$, which arise from the higher loops (or the two loops, at least), as above. 

Thus, it follows from our results that the two-loop corrected Starobinsky model, contrary to the one-loop corrected one, 
can provide the enhancement of the tensor-to-scalar ratio $r$ to the BICEP2 value.

Finally, it is worth mentioning that the considerations of this paper allow an extension to the N=1 supergravity, by using
curved superspace --- see e.g., Refs.~\cite{kw,kt}.

\section*{Acknowledgements}

This work was supported by a Grant-in-Aid of the Japanese Society for Promotion of Science (JSPS) under No.~26400252, the World Premier International Research Center Initiative (WPI Initiative), MEXT, Japan, and
the Competitiveness Enhancement Program of the Tomsk Polytechnic University in Russia. SVK is also
grateful to the DESY Theory Group in Hamburg, Germany, for kind hospitality extended to him during preparation of this paper, and M. Galante and M. Rinaldi for discussions and correspondence.


\begin{thebibliography}{9}

\bibitem{svrev} T. P. Sotiriou and V. Varaoni, Rev. Mod. Phys. {\bf 82} (2010) 451; arXiv:0805.1726 [gr-qc].
\bibitem{tsu} A. De Felice and S. Tsujikawa, Living Rev. Rel. {\bf 13} (2010) 3, arXiv:1002.4928 [hep-th]. 
\bibitem{norev} S. Nojiri and S. D. Odintsov, Phys. Repts. {\bf 505} (2011) 59; arXiv:1011.0544 [gr-qc].
\bibitem{clrev} S. Capozziello and M. De Laurentis, Phys. Repts. {\bf 509} (2011) 167; arXiv:1108.6266 [gr-qc].
\bibitem{myrev} S.V. Ketov, Int. J. Mod. Phys. {\bf A28} (2013) 1330021, arXiv:1201.2239 [hep-th].
\bibitem{star1}   A.~A.~Starobinsky,  Phys.\ Lett.\ B {\bf 91} (1980) 99.
\bibitem{star2} A.~A. Starobinsky,  {\it Nonsingular model of the Universe with the quantum 
gravitational de Sitter stage and its observational consequences}, in the Proceedings of the 
2nd Intern. Seminar on �Quantum Theory of Gravity�, Moscow, 13�15 October 1981 (INR Press,
Moscow, 1982), p. 58; reprinted in "Quantum Gravity", eds. M.~A. Markov and P.~C. West 
(Plenum Publ. Co., New York, 1984), p. 103.
\bibitem{mchi}  V.~F. Mukhanov and G.~V. Chibisov, JETP Lett. {\bf 33} (1981) 532.
\bibitem{star3}   A.~A.~Starobinsky, Sov. Astron. Lett.  {\bf 9} (1983)  302.
\bibitem{planck2} P.~A.~R.~Ade {\it et al.}  [Planck Collaboration],
 {\it Planck 2013 results. XXII. Constraints on inflation},  arXiv:1303.5082 [astro-ph.CO].
\bibitem{bicep2}  P.~A.~R.~Ade {\it et al.}  [BICEP2 Collaboration], {\it BICEP2 I: Detection of B-mode polarization at degree angular scales},  Phys. Rev. Lett. {\bf 112} (2014)  241101, arXiv:1403.3985 [astro-ph.CO].
\bibitem{bicep2c} R.~Adam {\it et al.}  [Planck Collaboration], {\it Planck intermediate results. XXX. The angular power spectrum of polarized dust emission at intermediate and high Galactic latitudes}, arXiv:1409.5738 [astro-ph.CO].
\bibitem{lwtr} B.~Whitt, Phys. Lett. {\bf B145} (1984) 17.
\bibitem{kkw} S. Kaneda, S. V. Ketov and N. Watanabe, Mod. Phys. Lett. {\bf A25} (2010) 2753, 
arXiv:1001.5118 [hep-th].
\bibitem{linde}  A. D. Linde, Phys.\ Lett.\ B {\bf 129} (1983) 177.
\bibitem{kwl} S.V. Ketov and N. Watanabe, Mod. Phys. Lett. {\bf A29} (2014)  1450117, arXiv:1401.7756.
\bibitem{stargu} V. T. Gurevich and A. A. Starobinsky, Zh. Eksp. Theor. Fiz. {\bf 77} (1979) 1683 (in Russian);
English translation in Sov. Phys. JETP  {\bf 50} (1979) 844. 
\bibitem{oe} E.~Elizalde and S.~D.~Odintsov, Phys. Lett. {\bf  B303} (1993) 240, arXiv:hep-th/9302074.
\bibitem{mass} M.~Rinaldi, G.~Cognola, L.~Vanzo and S. Zerbini, {\it Inflation in scale-invariant theories of
gravity}, arXiv:1410.0631 [gr-qc].
\bibitem{ijtwz} Ido Ben-Dayan, S. Jing, M. Torabian, A. Westphal and L. Zarate, {\it $R^2\log R$ quantum corrections and the inflationary observables}, arXiv:1404.7349 [hep-th].
\bibitem{kw} S.V. Ketov and N. Watanabe, Mod. Phys. Lett. {\bf A27}  (2012) 1250225, arXiv:1206.0416.
\bibitem{kt} S.V. Ketov and T. Terada, JHEP 1312 (2013) 040, arXiv:1309.7494 [hep-th].

\end{thebibliography}
\end{document}